\documentclass[twocolumns,letterpaper]{aa}
\usepackage{graphicx}
\usepackage{txfonts}
\begin{document}
   \title{Full radiative coupling in two-phase models for~accreting~black~holes}

\titlerunning{Full radiative coupling in two-phase models}
\authorrunning{J. Malzac et al.}
   \author{J. Malzac\inst{1,2} \and A.M. Dumont\inst{3} \and M. Mouchet\inst{3}}

   \offprints{J. Malzac, malzac@cesr.fr}

   \institute{Centre d'Etude Spatiale des Rayonnements, CNRS-UPS, 9
Avenue du Colonel Roche, F-31028 Toulouse Cedex 4, France\\
       \and
Institute of Astronomy, Madingley Road, Cambridge, CB3 0HA, United Kingdom\\          \and
LUTH, UMR 8102 (CNRS/Universit\'e 
Paris 7), Observatoire de Paris, Section de Meudon, F-92195 Meudon Cedex, 
France\\}

   \date{Received ??; accepted ??}

\abstract{
   The emission from galactic black holes and Seyfert galaxies
is generally understood in term of two-phase models
 (Haardt \& Maraschi 1991, 1993).
 Such models postulate that 
 a hot plasma ($\sim$10$^{9}$\,K) coexists with relatively 
colder material ($\sim$ 10$^{6}$\,K) in the inner part of the accretion flow.
We present the first simulated broad-band spectra
 produced by  such a system and accounting simultaneously for energy balance
 and Comptonisation in the hot phase, together with reflection, reprocessing,
ionization and thermal balance in the cold phase.
This was made possible by coupling three radiative transfer codes:
a non-linear Monte-Carlo code (NLMC), a photo-ionization code TITAN and
 a linear Monte-Carlo code NOAR. The equilibrium comptonisation
spectrum appears to be sensitive to the shape of the reprocessed
spectrum that, in turn, depends on the ionization parameter, but also
 on the structure of the irradiated cold 
material. This is illustrated by a comparison of simulations assuming
 constant density or a constant pressure in the cold phase.
 We also compare our results with simplified 
models where reprocessing is approximated by a blackbody spectrum.
Our detailed treatment leads to noticeably different spectral energy distributions (SEDs) characterised
by harder X-ray spectra. { Even at low ionization parameters
 ($\xi \sim 300$ erg~s$^{-1}$~cm)
the commonly used blackbody approximation is poor,
leading to X-ray spectra that are too soft}.
The effect, however, seems not to be strong enough to reconcile the slab 
corona model with the hardest observed spectra, { unless the
reflector has a constant density and the ionization parameter is
large, of the order of 10$^{4}$ erg~s$^{-1}$~cm.}

}

\maketitle

\section{Introduction}

At least two distinct mediums are required to produce 
the main features observed in the broad band spectra of radio quiet active
 galactic nuclei (AGN) and  black hole binaries (BHBs) in the low/hard state.
 Their hard X-ray  emission exhibits a power-law spectrum cutting-off
 at a few hundred keV, which is generally interpreted 
as thermal Comptonisation spectra in a very hot ($\sim$10$^{9}$ K)
 optically thin plasma with Thomson depth $\tau_{\rm T}\sim$1
(Sunyaev \& Titarchuk 1980; {  Haardt \& Maraschi 1991; Stern et
al. 1995b; Poutanen \& Svensson 1996; Zdziarski et al. 1998)}.
The big blue bump observed in the UV spectra of Seyfert galaxies (Walter 
\& Fink 1993),
 as well as the soft X-ray excess in BHBs (e.g. Balucinska-Church et al. 1995) are indicative of thermal emission from optically thick and much 
colder material with a temperature in the range  10$^{5}$--10$^{7}$ K.
The presence of relatively cold material in the innermost part of the 
accretion flow is corroborated by the existence of reflection features
 in the X-rays such as the Fe fluorescence line around 6.4 keV 
and a reflection bump peaking at about 30 keV 
(George \& Fabian 1991, Nandra \& Pounds 1994).

The nature and geometry of these two phases are uncertain.
The cold medium is generally believed to form
 an accretion disc (Shakura \& Sunyaev 1973), alternatively 
it could consist of cold dense clouds
located inside or around the hot plasma 
(Rees 1987, Collin-Souffrin et al. 1996, Malzac 2001, Malzac \&
Celotti 2002).
The hot phase could constitute the accretion disc corona (
Svensson \& Zdziarski  1994, \.Zycki et al. 1995, Sincell \& Krolik 1997,
R{\'o}{\.z}a{\'n}ska \& Czerny 2000, and ref. therein) or 
the hot inner part of the accretion disc itself (Shapiro \& Lightman 1976). 

It was soon realized that, if close to each other,
 the hot and the cold phases should be strongly radiatively coupled.
This led to the development of two-phase models (Haardt \& Maraschi 1991, 
1993), where the cold phase constitutes the main source of seed photons for 
the Comptonisation process. The temperature in the hot phase, and thus the
 detailed shape of the high energy spectrum, is  
controlled mainly by the flux of soft cooling photons. 
On the other hand, the thermal reprocessing of the high energy radiation
 impinging on the cold phase provides a dominant contribution to soft 
emission from the cold phase.

Due to the complexity of the Comptonisation process, the importance
 of geometric effects, and the requirement of taking accurately into 
account the coupling between the two phases, it was necessary to develop
 sophisticated numerical codes in order to compute the detailed
 spectrum produced by the two-phase system in different configurations
 (Stern et al. 1995a, Poutanen \& Svensson 1996).

The methods developed in this context, such as the non-linear Monte-Carlo
 method (Stern et al. 1995a), provide an accurate treatment of 
the hot phase emission in energy balance. On the other hand, the
 emission from the cold phase is not detailed: a pure blackbody 
spectrum with a fixed temperature and reflection on neutral material are
generally assumed.

Actually the emission due to X-ray reprocessing in the cold
 phase differs widely from a simple blackbody, in particular
 there is a complex line emission. 
Moreover, due to the hard X-ray irradiation, a ionised skin 
is likely to form on the surface layers of the cold material 
(Ross \& Fabian 1993, Collin-Souffrin et al. 1996, Nayakshin et al. 2000, 
Ballantyne et al. 2001). This ionised material affects 
the line emission as well as the shape and amplitude of the reflection
 component (\.Zycki et al. 1994, Dumont et al. 2000, Nayakshin et al. 2000, Nayakshin \& Kallman 2001). Ionization is also 
likely to affect the equilibrium in the hot phase. Indeed the hard X-ray
 albedo for ionised material is larger and, as a consequence,
 the flux of thermal reprocessed soft photons is lower, affecting
 the temperature in the hot plasma.

The first attempts to include these effects in the two-phase model
calculations used a very simplified approach to the computation of the
reprocessed spectrum  (e.g. Nayakshin \& Dove 2001).
Indeed, a detailed self-consistent computation of the emission from 
irradiated material is an heavy task. It requires one to solve the
 radiative transfer equations in the cold medium taking into account
 the energy and ionization balance including
 all the atomic processes. 
Another difficulty comes from the fact that the cold material is optically 
thick and its  temperature and ionization structure at equilibrium
 is strongly inhomogeneous.

The detailed computation of the structure and spectrum of irradiated 
optically thick material was performed by coupling  the photo-ionization
 code TITAN  and the Monte-Carlo code NOAR  (accounting for Compton 
scattering) (Dumont et al. 2000).
It enabled one to show the importance of an accurate radiative transfer treatment
 for the irradiated cold material emission.

However, in these calculations the feedback from the cold radiation on 
the hot phase equilibrium was not considered. As irradiating spectrum,
 a simple power-law with a fixed slope was assumed.

Here we add a self-consistent treatment of the hot phase emission
by coupling the TITAN/NOAR codes with a third code based on the 
Non-Linear Monte-Carlo method (Malzac \& Jourdain 2000).

This enables us to propose the first full treatment of the radiative
 coupling, accounting accurately and self--consistently
 for both the hot and cold phase emission. 

 \begin{figure}
  
   \includegraphics[width=\columnwidth]{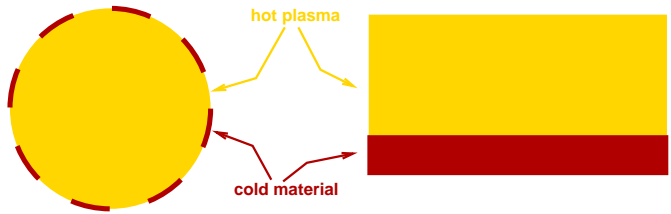}
  
\caption{The infinite slab accretion disc corona and spherical cloud
geometry. The hot slab is divided into 10 horizontal layers with
homogeneous temperatures. The sphere is divided into 5 concentric
shells.}  
\label{jmalzac1-C2_fig:fig1}
\end{figure}

 \begin{figure*}[!h]

\includegraphics[width=\textwidth]{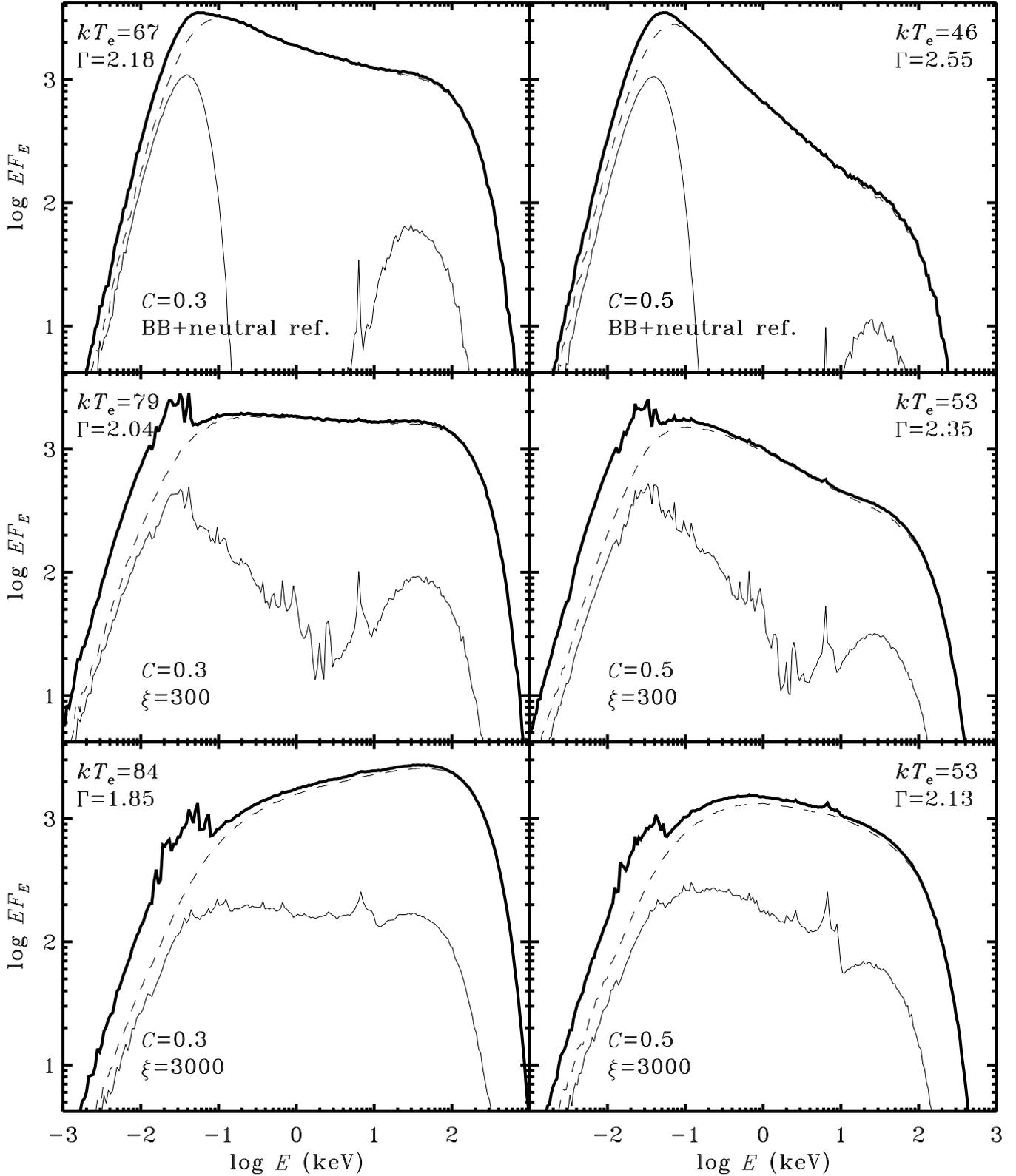}

\caption{Angle averaged escaping spectra for  covering fractions $C=0.3$ (left-hand side) and $C=0.5$ (right-hand side). Thin curves: reprocessed/reflected spectrum. Dashes: comptonised spectrum. 
Thick curves: total observed spectrum (including the transmitted component).
 The upper panels show the results from NLMC calculations with the cold phase emission approximated by a blackbody + neutral reflection. The blackbody temperature is $kT_{r}$=10 eV.
The other panels are the results from the TITAN/NOAR/NLMC calculations
for the following parameters of the cold medium:
$n\,=\,3\,10^{14}\,$cm$^{-3}$, $N_{H}\,=3\,10^{25}$\,cm$^{-2}$
($\tau_c=24$).
 In all of the simulations the Thomson optical depth of the hot plasma
was fixed to $\tau_{\rm T}=1$. The assumed ionization parameter $\xi$ is indicated in each panel together with the resulting 2-10 keV photon index of the comptonised emission $\Gamma$ and the volume averaged temperature of the hot phase $kT_{\rm e}$ (in keV).} 
\label{jmalzac1-C2_fig:fig2}
\end{figure*}

 \begin{figure*}[!h]

\includegraphics[width=\textwidth]{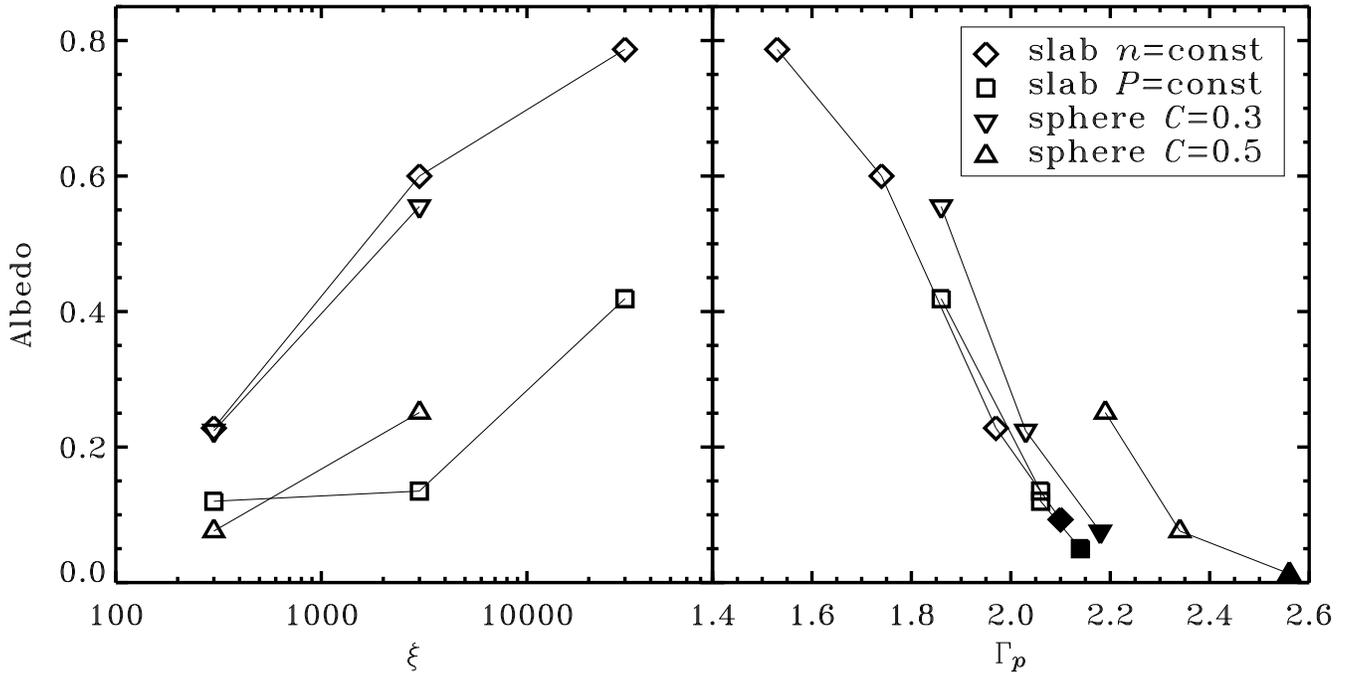}

\caption{X-ray albedo (i.e. fraction of the reflected/reprocessed
luminosity emitted above 1 keV) as a function of the ionization
parameter $\xi$ (left) and spectral index,  $\Gamma_p$ (right), of the
simulated spectra obtained from fits with the
PEXRAV model (see text and Table~\ref{tab:fits}). The results are
plotted using a different symbol for each model, as shown on the
figure. Filled symbols correspond to the blackbody and neutral
reflection models.}
\label{fig:albedo}
\end{figure*}

 \begin{figure*}[!h]

\includegraphics[width=\textwidth]{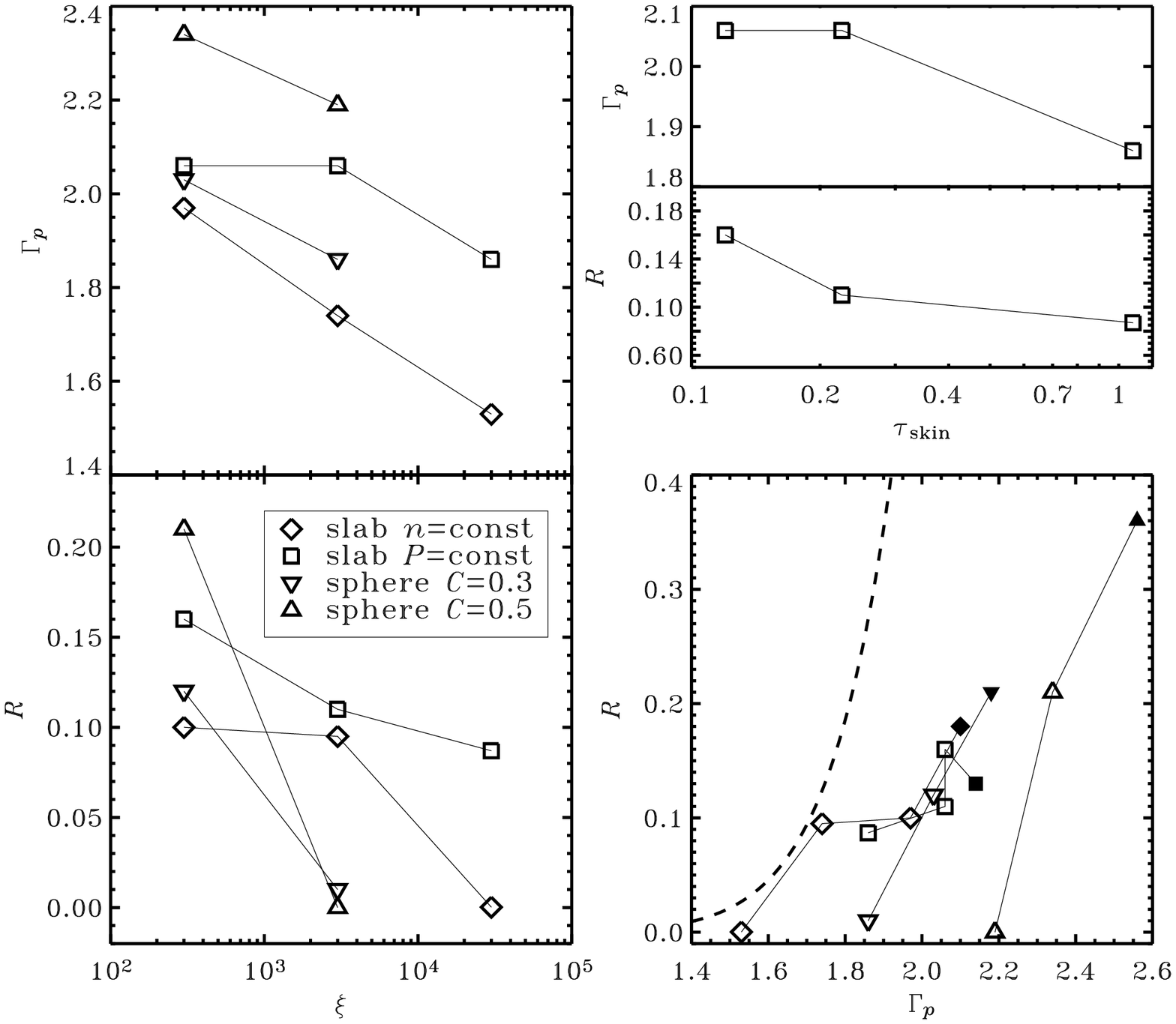}

\caption{Reflection amplitude $R$ and spectral index $\Gamma_p$ of the
simulated spectra obtained from fits with the
PEXRAV model (see text and Table~\ref{tab:fits}). The results are
plotted using a different symbol for each model, as shown in the
figure. Filled symbols correspond to the blackbody and neutral
reflection models. Left panels: $R$ and
$\Gamma$ as a function of $\xi$. Upper right panel: in the constant
pressure case, $R$ and $\Gamma_p$ as a function of the hot skin Thomson optical depth
$\tau_{\rm skin}$. Bottom right panel: $R$ versus $\Gamma_p$, the dashed
curve shows the best fit power law approximation to the observed $R$-$\Gamma$
correlation in AGN as given by Zdziarski et al. 1999. }
\label{fig:fits}
\end{figure*}

\section{The numerical model}

\subsection{The NOAR/TITAN and NLMC codes}

TITAN is a code designed for warm media (T$>$ a few 10$^{4}$ K) optically thick
to Compton scattering.
It computes the structure of a plane-parallel slab of gas in thermal
and ionization equilibrium, illuminated by a given spectrum
on one or two sides of the slab (Dumont et al. 2000).
It takes into account the returning flux using a two-stream approximation
to solve the transfer in the lines (instead of the escape probability
formalism). This code is coupled with a  Monte Carlo  code, NOAR, which
takes into account Compton and inverse Compton diffusions
in any geometry (Abrassart 2000). NOAR uses the local fractional ion abundances
and the temperature provided by TITAN, while NOAR provides
TITAN with the local Compton gains and losses in each layer.
The  Compton heating-cooling rate is indeed  dominated by energy losses
of
photons at high energies ($>$ 25 keV), not considered by TITAN.
The coupling thus allows one to solve consistently both the global and the local
energy balance.
NOAR also allows computing the fluorescence line profiles which are
significantly Compton-broadened in the case of strong illumination, and the 
Comptonised reflection spectrum above 25 keV.

In parallel, we use the NLMC code described and tested 
in Malzac \& Jourdain (2000) based on the Non-Linear Monte-Carlo 
method proposed by Stern et al. (1995a).
This code computes the Comptonised spectrum and energy balance
 in the hot phase.
The radiation field is represented using about 10$^{4}$
 pseudo-particles called Large Particles (LP). Each LP represents a number
 of photons with identical characteristics (energy, position, direction of
 propagation,...). These LPs are tracked all together in a synchronized way.
 They may interact by the Compton effect with a Maxwellian electron distribution.
These interactions are simulated using standard Monte-Carlo methods.
For a fixed heating rate in the hot phase, 
the temperature of the electron distribution is modified according to the 
Compton energy losses due to the interactions with the LP photons.
Starting from an arbitrary radiation field and temperature 
the system evolves naturally toward equilibrium. Then, the escaping LP
 characteristics are used to build up the angle dependent spectrum from 
the hot phase
 until satisfying photon statistics are achieved.

\subsection{Geometry and model parameters}

The proposed method can apply to a variety of geometries of the two-phase
system. { However as the coupled TITAN/NOAR/NLMC simulations are
very time-consuming, we will limit this first 
attempt of full radiative coupling to simple situations}. 
We will consider both the  slab and spherical geometry
illustrated in Fig.~\ref{jmalzac1-C2_fig:fig1}. 
The slab geometry,  used 
e.g. by Malzac \& Jourdain (2000), corresponds 
to the standard infinite slab accretion
disc corona model (Haardt \& Maraschi 1993). 
The spherical geometry corresponds to the model of 
Collin-Souffrin et al. (1996) where the hot phase constitutes a sphere 
at the center of the accretion flow surrounded by cold material
possibly in the form of dense clouds or filaments. For simplicity, we
assume that the material is homogeneously distributed at the sphere. 
For both models, the hot plasma is assumed to have an homogeneous density. 
It is however divided into a number of zones to account
 for temperature gradients that are
computed according to the local energy balance.

Our model is then fully determined when the following parameters are
given:\\
$\bullet$ Thomson optical depth $\tau_{\rm T}$ of the hot phase,
 defined along the radius of the sphere or the height of the hot slab.\\
$\bullet$  Density of the cold (homogeneous) material  $n$\,(cm$^{-3})$.\\
$\bullet$  Hydrogen column density of the cold
material\,$N_{H}$\,(cm$^{-2}$), or equivalently its Thomson optical
depth $\tau_{c}$.\\
$\bullet$ Ionization parameter of the cold material $\xi$, 
defined as $\xi=4\pi F_{bol}/n$, where $F_{bol}$ 
is the integrated flux incident on the cold material surface.
In the following $\xi$ is expressed in erg~s$^{-1}$~cm.\\
$\bullet$  Covering factor $C$ of the cold material, 
in the case of the spherical geometry. 
$C$ is the surface ratio covered by the
cold clouds to that of the sphere. For a photon escaping from the hot 
sphere, $C$ represents the probability of entering into the cold medium.\\

\subsection{Numerical method}

We compute the escaping spectrum as follows.
First, we use the non-linear Monte-Carlo code in order to get a 
first estimate of the high energy spectrum.
When a LP photon escapes the hot plasma and enters the cold material,
its energy is then reemitted at its surface in the form of
reprocessed/reflected LP photons.
 Their energy and direction are drawn from the assumed angle dependent
 spectrum of the cold material. Note that the reprocessed photons may
be either directed toward the hot plasma (i.e. reflected) or directed
outward and escape the system (i.e. transmitted through the cold phase).\\

In the spherical model, we use a statistical method to model 
the homogeneous distribution of cold material around the
sphere.
When a photon LP escapes the hot sphere, a random number $\eta$ is drawn:\\
$\bullet$ if $\eta > C$,  the LP photon truly escapes and its energy is used to build up the escaping spectrum.\\
$\bullet$ if $\eta < C$,  the LP photon enters the cold material and
its energy is either reflected toward the sphere 
or transmitted as described above.

At this stage the cold medium spectrum is arbitrary
 (although it is better if it is similar to a real reprocessed spectrum).
We then inject the resulting equilibrium hot-phase spectrum as input 
in the TITAN/NOAR codes.  
This provides an estimate of the ionization and temperature structure 
of the cold phase
 as well as the reprocessed spectrum. 
Then we use the TITAN/ NOAR spectrum as the local spectrum 
of the cold material in the NLMC code to get a better estimate of
 the hot-phase spectrum, and so on. 
In general, convergence is achieved after  3-4 iterations.

\begin{figure*}[h]

\includegraphics[width=\textwidth]{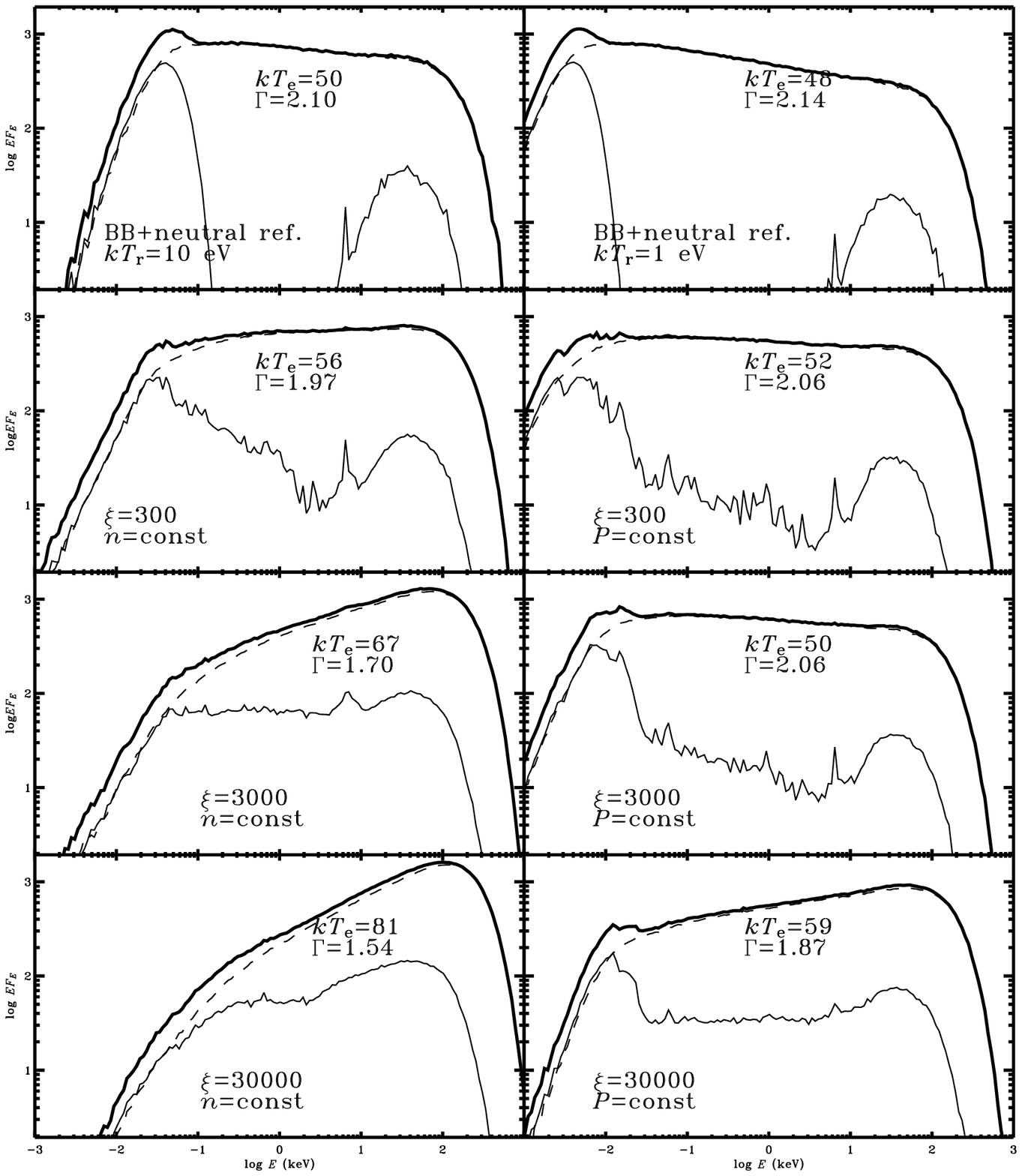}

\caption{Angle averaged escaping spectra for a plane-parallel geometry.
The bottom panels are the results from the TITAN/NOAR/NLMC calculations
for the same parameters of the cold medium as in Fig.~\ref{jmalzac1-C2_fig:fig2}, and
assuming a constant density (left hand side) and a constant
pressure (right hand side) in the cold phase.
 The upper panels show the results from NLMC calculations with the
cold phase emission approximated by a blackbody + neutral
reflection for two different blackbody temperatures $kT_{r}$=10 eV
(left hand side)  $kT_{r}$=1 eV (right hand side).
 Thin curves: reprocessed/reflected spectrum. Dashes: Comptonised spectrum. 
Thick curves: total observed spectrum.
In all panels, the Thomson optical depth of the hot plasma is $\tau_{\rm T}=1$.The assumed ionization parameter $\xi$ is indicated in each panel together with the resulting 2-10 keV photon index of the Comptonised emission $\Gamma$ and the volume averaged temperature of the hot phase $kT_{\rm e}$
 (in keV).} 
\label{fig:slabncst}
\end{figure*}

\section{Results}

\subsection{Spherical model} \label{sec:sphmo}

Figure~\ref{jmalzac1-C2_fig:fig2}  shows the resulting spectra for
$C$=0.3 and $C$=0.5 and two values of the ionization parameter 
 $\xi=300$ and $\xi=3000$ assuming a constant density in the cold phase.  
 
For comparison we also show spectra that were obtained with
 the usual treatment of reprocessing 
(i.e. blackbody spectrum + neutral reflection). 
If all the radiation impinging on the cold material is absorbed
and reemited as a blackbody, its temperature
is related to the incident flux through 
the Stefan-Boltzmann law. Therefore the temperature  $kT_{r}$ of the
cold phase is related to the ionization parameter and density
 of the cold phase as follows:
\begin{equation}
T_{r}=\left(\frac{\xi n}{4\pi\sigma}\right)^{1/4},
\end{equation}
where $\sigma$ is the Stefan's constant.
As a result, this equivalent blackbody temperature is not
very sensitive to the incident flux.
For our assumed density, $kT_{r}$ is 7 and 12 eV for
$\xi$=300 and 3000 respectively.
In accordance, in the blackbody spectrum + neutral reflection simulations, 
the blackbody temperature was fixed 
at $kT_{r}$=10 eV in both cases, all the other relevant parameters
 being kept at the same values.
Indeed, one can see from Figure~\ref{jmalzac1-C2_fig:fig2} that the blackbody
spectrum peaks at nearly the same energy as the TITAN/NOAR spectrum.
 
Globally, for larger $C$, the X-ray spectra
are always softer, i.e. the photon index of the primary emission
$\Gamma$ (estimated with a least square fit of the 2-10 keV primary spectrum) is larger. This is due to the increase of reprocessed
cooling flux from the cold material at larger $C$ which leads to a lower temperature in the hot-phase.

There are however noticeable differences  between spectra obtained
with the usual treatment (i.e blackbody spectrum + neutral reflection) 
and the detailed treatment. The shape of the reprocessed spectrum obviously differs. But the Comptonised
 emission differs as well. In particular, the spectrum is much harder in the calculations using TITAN/NOAR.
Indeed the reprocessed spectrum then 
is much broader than a simple blackbody, more photons being reprocessed 
at higher energy. This affects the Compton cooling in the hot phase (which is lower) together with the shape of the Comptonised spectrum which 
is harder also due to the relatively high energy of the seed photons. 
This effect is then amplified by the fact that the hard X-ray albedo is larger
for a harder spectrum thus enhancing the fraction of the reflected/reprocessed photons at high energy.
We note that in the ``blackbody'' simulations, the total flux impinging 
on the cold material is reemitted toward the hot sphere
 (i.e. transmission through the cold medium is neglected).
 In contrast, in the TITAN/NOAR simulations,
 the transmitted flux\footnote{By `transmission' or `transmitted flux' we
refer to the net energy flux across the cloud. This includes the
radiation crossing the cloud without interaction, but also the reprocessed
radiation that is outwardly reemitted. The latter is largely dominant.}

 represents about 15 \%  of the impinging luminosity,
 despite our choice of a relatively large hydrogen column density of the 
cold medium { (corresponding to a Thomson optical depth
$\tau_c$=24).} 

 This modest ``loss'' of  soft photons leads to an increase of the
temperature of a few percent and a hardening of the spectrum by
$\Delta \Gamma \simeq 0.03$. 
Therefore, this effect does not 
contribute much to the resulting  higher temperatures and  harder
spectra in the TITAN/NOAR case.

The hardening of the spectrum increases with $\xi$
 (compare the spectra for $\xi=3000$ and $\xi=300$).
 The strongly ionised material has indeed a larger 
X-ray albedo, and more photons are reprocessed at high energies.
We note that the differences between the blackbody + neutral reflection model 
and the realistic one appear to be important even at moderate
 $\xi\sim300$.  The left panel of Fig.~\ref{fig:albedo} 
shows the dependence of the
X-ray albedo (defined as the fraction of the reflected/reprocessed
luminosity emitted above 1 keV) on $\xi$. As the X-ray albedo also strongly
depends on the shape of the incident spectrum (see Malzac et
al. 2001), for a given $\xi$ the albedo can vary significantly  
with the  geometry considered. The models with 
$C=0.3$ and $C=0.5$ have albedos differing by more than a factor of 2,
that is as important as the change in albedo when $\xi$ is increased
by a factor of 10. The right panel of Fig.~\ref{fig:albedo} shows the
resulting relation between albedo and spectral index. It is
interesting to note that the difference of the albedos for the neutral 
and the low ionisation models ($\xi=300$) is too low to account for
the important changes in $\Gamma$ that we obtain. 
Indeed, the albedo changes by  about 0.1, implying an increase in the
soft cooling flux of about 10 \%, which, as noted above, is not
sufficient to increase $\Gamma$ by $\Delta \Gamma \simeq 0.2$.
This shows that the important difference in spectral slope between
neutral and weakly ionised cases is essentially due to the different
shape of the reprocessed spectrum below 1 keV (as explained
above) and not much to changes in the X-ray albedo.

   \begin{table}[b]
      \caption[]{Spectral parameters obtained when fitting our
simulated spectra in the 2--30 keV range under XSPEC 
with the PEXRAV model plus a Gaussian line. 
$\Gamma_{p}$ and $R$ are the PEXRAV photon index and reflection
amplitude. $E_{\alpha}$ is the line centroid energy in keV, and EW is its
equivalent width in eV.  { During the fit the high energy cut-off energy
$E_{c}$ and the inclination angle  were fixed at 400 keV and
30~degrees respectively, and the abundances were fixed to standard.}}

\begin{tabular*}{0.99\columnwidth}{@{\extracolsep{\fill}}llccccc}

           \hline
            \noalign{\smallskip}
             simulation  & & $\Gamma_{p}$  & $R$ &   $E_{\alpha}$ & EW\\
            \noalign{\smallskip}
            \hline
            \noalign{\smallskip}
C=0.3 &  $T_r$=10 eV  & 2.18  &    0.21 & 6.40 &  3.12 \\
C=0.3 & $\xi$=300    & 2.03  &    0.12 & 6.32 &  29.8  \\
C=0.3 & $\xi$=3000   & 1.86  &    0.01 & 6.61 &  126  \\
C=0.5 & $T_r$=10 eV  & 2.56  &    0.36 & 6.40  &  8.66 \\
C=0.5 & $\xi$=300    & 2.34  &    0.21 & 6.31 &  35.6\\
C=0.5 & $\xi$=3000   & 2.19  &    0.00 & 6.41 &  451   \\
       \noalign{\smallskip}
       \hline
        \noalign{\smallskip}
slab  &  $T_r$=10 eV  &  2.10 &  0.18 &   6.40  &    8.0   \\

slab  $n$=const &  $\xi$=300     &  1.97 &  0.10 &   6.56  &    53.5
\\
slab  $n$=const &  $\xi$=3000     &  1.74 &  9.5 $10^{-2}$ &   6.1  &    307
\\
slab  $n$=const &  $\xi$=30000     &  1.53 &  2.1 $10^{-4}$ &   1.32  &  32  
\\
slab  &  $T_r$=1 eV   &  2.14 &  0.13 &   6.40  &    9.53 \\
slab  $P$=const &  $\xi$=300    &  2.06 &  0.16 &   6.4  &    12.9 \\
slab  $P$=const &  $\xi$=3000    &  2.06 &  0.11 &  6.4  &    18.8 \\
slab  $P$=const &  $\xi$=30000    &  1.86 &  8.7 $10^{-2}$ &  6.4  &    7.45 \\
         \hline
         \end{tabular*}
\label{tab:fits}
   \end{table}

Besides the spectral index, there are other parameters, such as the
amplitude of the reflection features, that may be relevant to the observations.
To derive those parameters from the simulated spectra in a way similar
to those from  observed spectra, we fitted our simulated
spectra using the X-ray spectral fitting package { XSPEC} (Arnaud, 1996). The fitting model
was an e-folded power law plus neutral reflection ({PEXRAV}, Magdziarz
\& Zdziarski 1995) and a Gaussian to model the iron line.
The results are summarised in Table~\ref{tab:fits} and
Figure~\ref{fig:fits}. In all cases we found that the spectral index
obtained with {PEXRAV}  $\Gamma_{p}$, is very close to that derived
through the 2-10 keV least square fit.
$\Gamma$. They show that the
reflection amplitude $R$, measured with {PEXRAV}, strongly depends on the
ionization parameter. It decreases by almost a factor of two between
the neutral and $\xi=300$ case and for the highest ionization
parameter $R$ is almost 0. This is a well known effect due to the fact 
that the more ionised the reflector is, the more the reflected X-ray
spectrum is similar to the incident one due to the predominance of
Compton reflection over photoabsorption. 
As a consequence, the reflection bump
appears relatively weaker in the overall spectrum.
A further reduction of $R$ is due the destruction of the reflection
component by Comptonisation in the hot phase (Malzac, Beloborodov \&
Poutanen 2001; Petrucci et al. 2001) depending on the Thomson optical
depth of the hot sphere.

Contrary to the reflection component, the equivalent width of the iron
line increases with the ionization parameter. Indeed in our
simulations the line intensity increases with $\xi$ (by a factor of 10
between $\xi$=300 and $\xi=3000$), This is in agreement with previous
studies showing that the line intensity increases between these two
values of $\xi$ we considered  (Matt, Fabian \& Ross 1993, 1996). We
note that in our fits the line energy is very weakly constrained. This
is due to the low resolution of the fitted simulated spectra and also
probably to the inadequacy of the Gaussian approximation to represent
the much more complex profile of the actual line. This may explain the
few unphysical red-shifted lines we obtain. Indeed, if estimated
directly from the simulation, the line peak is always above 6.4 keV 
and blue shifted at large $\xi$.

How do these results compare with the observations ?
The photon indices $\Gamma_{p}$$>$2 obtained with PEXRAV for most
 of our simulations correspond to the softest observed sources
(see e.g. Zdziarski et al. 2003).
In the context of the spherical model this suggests 
that either the cold material is usually strongly ionised,
 or that in most sources, the covering fraction is lower
than 0.3. On the other hand, the resulting $R$ is
rather low compared to the average $R\sim0.7$ in Seyfert galaxies,
suggesting on the contrary that in most sources the covering fraction is larger. 
These problems can be easily overcomed by considering a slightly
different geometry where the cold material, instead of being at the
sphere surface would be at some distance. This would decrease both the cooling of the hot phase and the fraction of reflected photons that are Comptonised. Then, as shown by Malzac (2001),
for the same covering fraction the spectrum is harder and the
reflection component is stronger. We also note that, in Seyfert galaxies,
a distant molecular torus could provide a strong additional  
contribution to the observed reflection component (Malzac \& Petrucci 2002).

\subsection{Slab corona model}

In the spherical model, changes in the covering fraction of cold
material may produce a range of X-ray photon indices that includes the observed
range. On the other hand the slab corona model, { in its simplest
version}, does not allow such
freedom; the photon index is tightly constrained 
 (Haardt \& Maraschi 1993). Detailed simulations using the
blackbody + neutral reflection approximation (Stern et al. 1995b)
 have shown that the 
range of $\Gamma$ expected in this geometry tends to be larger than
$\sim 2$, while the average $\Gamma$ in Seyfert 1 galaxies
is~$\sim1.8$. Observed spectra of Seyfert galaxies and galactic black
hole sources can be much harder with photon indices as low as $\Gamma$$\sim$1.4.
{ Overcoming this problem requires  some complications,
 such as considering the effects of an out-flowing corona 
(Beloborodov 1999; Malzac, Beloborodov \& Poutanen
2001) or the presence of holes in the disc  (Zdziarski et al. 1998).
As a consequence, the simple slab coronal model considered here
 was often considered as ruled out on this basis.} 

As our more realistic treatment of reprocessing and ionization
 tends to produce harder spectra, 
it is interesting to see whether this would be sufficient 
to reconcile the model with the observations.
The left panels of Fig.~\ref{fig:slabncst} compare spectra for a 
slab geometry and both  the blackbody approximation and realistic
reprocessing for $\xi=3\times10^{2}$, $\xi=3\times10^{3}$ and $\xi=3
\times 10^{4}$, assuming a constant density in the cold
phase. {  Unlike those of  the spherical model, these slab spectra are
angle dependent. However the weak angular dependence obtained in our
detailed reprocessing simulations does not differ qualitatively from
what is found in the standard  
'blackbody' +neutral reflection treatment that was already studied in
several papers (see e.g. Malzac, Beloborodov \&
Poutanen 2001). For the sake of simplicity,  we will consider the
properties of the angle-averaged spectra only, 
since the aim of this work is to evaluate
 the main effects of a detailed treatment of reprocessing and
ionization.}
   
The spectral indices from the 2-10 keV least square fit  are
$\Gamma$=2.10 for the 'blackbody' simulation  and
$\Gamma$=1.97, 1.70 and 1.54  respectively for the ionised cases.
 Thus although the hardening of the spectrum 
is significant, the model will  be
able to reproduce the spectra of the hardest sources only if the disc
is strongly ionised.

Since, in our calculation, we neglect the contribution of internal
viscous dissipation to the disc emission, our spectra are the hardest
possible for the slab corona geometry. Any intrinsic disc emission would make
the spectra softer and strengthen the requirements for extreme
ionization.

In the TITAN/NOAR simulations shown in Fig.~\ref{jmalzac1-C2_fig:fig2} 
and~the left panels of Fig.~\ref{fig:slabncst} we assumed a constant density in the
cold material. This assumption might be a reasonable approximation in
the case of the  small-scale magnetically confined clouds of the model of
Collin-Souffrin et al. (1996).
 In the case of the accretion disc corona however, 
there are certainly strong density gradients driven by 
the pressure equilibium.  
In general, due to the ionization thermal instability, it breaks into
two well defined layers: a low density skin at Compton
temperature that is almost completely ionised, atop the cooler, high
density, low ionization material of the internal disc.
    The sharp transition between the two layers contrasts with the smooth 
ionization gradients obtained in the constant density case.
This characteristics was studied in a number of papers 
 (Raymond 1993; Ko \& Kallman 1994; R\'{o}\.{z}a\'{n}ska \& Czerny 1996;
Nayakshin, Kazanas \& Kallman 2000; Dumont et al. 2002) that showed
it has strong impact on the reprocessed spectrum.
The reprocessed spectrum can then be described as formed by two 
components associated with two layers. 
The hot skin component is essentially dominated by Compton
reflection with a spectral shape almost indistinguishable from the
irradiating spectrum, while the deeper component corresponds to almost 
neutral reprocessing (Done \& Nayakshin 2001; Ballantyne, Ross \&
Fabian 2001).

Our results indicate that the primary spectrum is sensitive
to the shape of the reprocessed spectrum.
 Therefore, one may wonder what effect 
different assumptions made about the structure of the cold material (constant
density, constant pressure or hydrostatic equilibrium)
may have on the shape of the overall equilibrium spectral energy distribution.

In the right panels of Fig.~\ref{fig:slabncst}, we show the resulting spectra
 for the slab geometry and a constant pressure in the cold phase. { For the 3 simulations with constant pressure the resulting
Thomson depth of the ionised skin $\tau_{\rm skin}$ is 0.120,
0.225, and 1.08 respectively for $\xi=3\times 10^{2}$, $3\times 10^{3}$, and $3\times 10^{4}$}.
As compared to the constant density case, and in agreement with
previous studies, the reprocessed spectrum 
is much softer in the X-ray range due to the strong fraction of
photons  that are Compton reflected in the hot skin and the low
ionization state of the { deeper layers}. For the same reasons, the UV
spectrum formed in the colder deeper regions 
is sharper and peaks at lower frequency.
The soft seed photons are thus softer on average 
and as a consequence the primary X-ray 
spectrum is slightly steeper 
than its constant density counterpart.

Despite the reprocessed spectrum, in its lowest energy part,
 is now closer to a blackbody spectrum,
the primary X-ray emission still 
differs from that obtained in the blackbody approximation. 

To facilitate the comparison, the 'blackbody' case shown in
upper right panel of Fig.~\ref{fig:slabncst} is for $kT_{r}$=1 eV,
 so that the blackbody peaks nearly at the same energy
 as the realistic reprocessed UV spectrum.
For both  $kT_{r}$=1 eV and $kT_{r}$=10 eV the primary X-ray spectrum is
steeper than in the TITAN/NOAR simulation.

The results  of the XSPEC fitting procedure are shown in
Table~\ref{tab:fits}.
For a very thick reflector subtending a solid angle
2$\pi$, we expect $R=1$. On the other hand, the
reflection coefficient we derive are relatively low,
 $\sim$ 0.1--0.2 even in the
case of neutral reflection. { This is a known effect resulting} 
 from the destruction of  the reflection component 
by Comptonisation in the hot phase 
(Malzac, Beloborodov \& Poutanen 2001; Petrucci et al. 2001). In general, 
this effect is important for all geometries where the reflector
 is seen through the hot plasma, such as the slab or spherical
geometry considered in Sect.~\ref{sec:sphmo}.

 { Contrary to the case of a
constant density, in the constant pressure models, 
 the spectral index is quite insensitive to the
ionization parameter. The spectrum hardens by less than $\Delta \Gamma
\simeq 0.2$ when $\xi$ is increased by 2 orders of magnitude.}  
In this case the ionization effects are not sufficient, to 
reconcile the slab corona model with the observations.
This latter result confirms a similar conclusion
 reached by Nayakshin \& Dove (2001),  
on the basis of a simplified treatment of the hot skin model.
{ On the other hand, if the reflector has a constant density, 
we find that the different ionization parameters could explain the whole range
of observed spectral indices. This however would require a very wide range of 
ionization parameters, with an extremely ionised
reflector in the hardest sources. Moreover, although the constant
density is plausible in the context of the cloud model of
Sect.~\ref{sec:sphmo}, an accretion disc with a constant density
 seems quite unrealistic.}

Finally, both for constant density and constant pressure a
detailed treatment of reprocessing appears to be important.
 The usual blackbody+neutral reflection approximation  appears
 inappropriate for an accurate determination of the equilibrium
spectra, even at low ionization.

\section{Conclusions}

We calculated the equilibrium spectra in the context of the two-phase
 models for the emission from Seyfert galaxies and black hole binaries. 
For the first time, we included 
a detailed treatment of reprocessing and ionization and thermal balance 
in the cold phase.

The resulting broad band spectrum differs significantly from what obtained 
using the usual ``blackbody + neutral reflection'' approximation.
In particular, our realistic 
treatment of reprocessing leads to higher hot-phase temperatures
 associated with  harder X-ray spectra.
{  Surprisingly the effect is
strong even at low ionization parameters. This is because even if the X-ray
albedo is not significantly affected by the weak ionizing illumination,
the shape of the reprocessed spectrum below 1 keV is much broader than
a simple blackbody.}

For the slab model, we performed simulations with two different prescriptions 
for the cold phase, namely a constant density and a constant pressure. 
The comparisons between the two cases indicate that the 
overall spectral shape is also quite sensitive to 
the physical structure of the cold phase. A slab model
 at constant density can account for the hardest 
observed spectra for large values of the  ionization parameters.
In the case of constant pressure disc  however, the hardening 
is not strong enough to reconcile the model
 with the even harder spectra observed in  
Seyfert galaxies and Galactic black holes.

The equilibrium spectra of a two-phase system thus appear to be 
significantly affected by the shape of the reprocessed spectrum.
 even at low ionization parameters.
A detailed treatment of reprocessing appears to be
required for most observationally relevant cases.

\begin{acknowledgements}
 This work was partly supported by the European Commission
 (contract number ERBFMRX-CT98-0195, TMR network
"Accretion onto black holes, compact stars and protostars").
JM also acknowledges fundings from the MURST (COFIN98-02-15-41) and
PPARC. JM acknowledges a travel grant from the GDR PCHE.
We are grateful to Suzy Collin and Pierre-Olivier Petrucci 
for enlightening discussions.
\end{acknowledgements}

\end{document}